\documentclass[6pt, preprint2]{aastex}
\usepackage[]{natbib}
\begin{document}
\title{Intrinsic Redshifts in QSOs Near NGC 6212}
\author{M.B. Bell\altaffilmark{1} and S.P. Comeau\altaffilmark{1}}
\altaffiltext{1}{Herzberg Institute of Astrophysics,
National Research Council of Canada, 100 Sussex Drive, Ottawa,
ON, Canada K1A 0R6}

\begin{abstract}

The high number of QSOs around NGC 6212 allows a correlation analysis to be carried out between their redshift distribution and the redshift distributions predicted by intrinsic redshift models. We find no correlation between the QSO redshift distribution and the intrinsic redshifts predicted by the $\Delta$log(1+z) = 0.089 relation. However, we find that the QSO redshift distribution is correlated with the intrinsic redshifts predicted by the relation z$_{iQ}$ = 0.62[N-0.1M$_{N}$], for $N$ = 3. We also find evidence that the observed redshifts of the QSOs contain a small cosmological redshift component similar to that of NGC 6212. The same correlation analysis carried out on the distribution of 574 quasar redshifts discussed by Karlsson also gave a negative result for the $\Delta$log(1+z)= 0.089 relation, and a positive result for the z$_{iQ}$ intrinsic redshifts. 
\end{abstract}


\section{Introduction.}

The location of 41 QSOs near the Seyfert 1 galaxy NGC 6212 \citep{bur03} is shown in Fig 1. The QSOs have redshifts that range from 0.03 to 2.529. Even if the area outside this region of high density has not been surveyed with the same care and intensity for QSOs, this figure is still an impressive argument for the existence of intrinsic redshifts. If these objects have been ejected from NGC 6212 (z$_{o}$ = 0.03) as suggested by \citet{bur03}, their observed redshifts will be given by the relation

\begin{equation}
(1+z_{o}) = (1+z_{c})(1+z_{\rm D})(1+z_{i})
\end{equation}
where z$_{c}$ is the cosmological component of the observed NGC 6212 redshift, z$_{\rm D}$ is a Doppler component due to the l-o-s ejection velocity, and z$_{i}$ is the intrinsic component. Note that it is assumed here that the observed redshift of the Seyfert galaxy may also contain a small intrinsic redshift component. Thus, 

\begin{equation}
(1 +z_{o})_{NGC6212} = (1 + z_{iG})(1 + z_{c}) = 1.03
\end{equation}
where z$_{iG}$ is the intrinsic component of the galaxy, and z$_{c}$ is its cosmological or distance component. We have found evidence that the z$_{iG}$ redshifts in galaxies may correspond to velocities as high as several thousand km s$^{-1}$ \citep{bel02d,bel03a,bel03b,tif97}, implying that the latter (z$_{c}$) is likely to be less than z$_{o}$ = 0.03. This possibility needs to be kept in mind if small cosmological redshift corrections to the observed QSO redshifts are contemplated, since an intrinsic component present in the observed redshift of NGC 6212 will not be present in the QSO redshifts.

We assume for the purposes of this analysis, that most of the QSOs near NGC 6212 have been ejected from the Seyfert galaxy. If this is true, for a given intrinsic redshift model the biggest unknown in eqn 1 is likely to be z$_{\rm D}$. However, as discussed below, it may be possible to estimate this ejection velocity redshift. In what follows we first carry out a correlation analysis between the redshift distribution obtained for the QSOs near NGC 6212 and the intrinsic redshifts predicted by two intrinsic redshift models. We then carry out a similar analysis using the redshift distribution obtained for the 574 quasars discussed by \citet{kar71,kar77}.

\begin{figure}
\hspace{-1.0cm}
\vspace{-2.0cm}
\epsscale{1.0}
\plotone{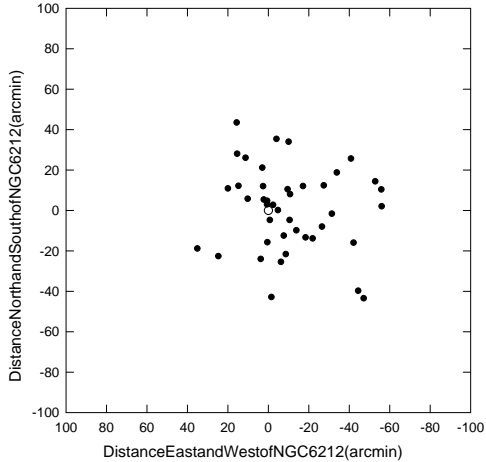}
\caption{\scriptsize{Plot of QSOs around NGC 6212. North is at the top and East to the left. \label{fig1}}}
\end{figure}

\begin{figure}
\hspace{-1.0cm}
\vspace{-2.0cm}
\epsscale{1.0}
\plotone{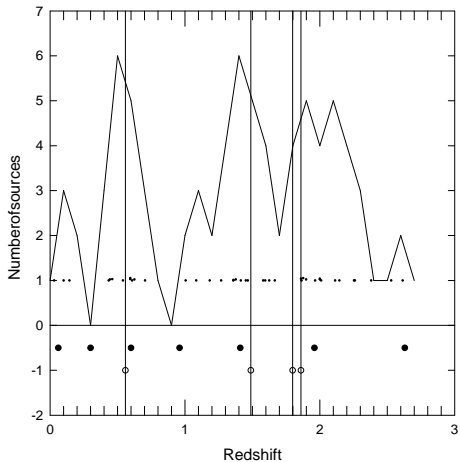}
\caption{\scriptsize{Distribution of redshifts in QSOs near NGC 6212. The three broad peaks in the distribution curve indicate the regions where the density is highest. Filled circles are redshifts predicted from the Karlsson relation. Open circles correspond to intrinsic redshifts from the z$_{iQ}$ = 0.62[$N - 0.1 M_{N}$] relation for $N$ = 3. \label{fig2}}}
\end{figure}

\section{Analysis}

The redshift distribution for the QSOs near NGC 6212 is shown in Fig 2, plotted at level 1. Also included in the figure is a curve obtained by taking a running mean over the redshift data with a bin width of z = 0.2. There are three redshift regions that appear to be favored: z$_{o} \sim 0.52, \sim 1.5$ and $\sim 2$. If the redshift spread in each of these three redshift clumps is due to smearing by l-o-s components of the ejection velocities of these objects, it can be explained nicely by ejection velocities near 20,000 km s$^{-1}$ (z$_{\rm D} \sim0.07$).

\subsection{Correlation between observed redshifts and those predicted by intrinsic redshift relations}

\citet{kar71,kar77} has claimed that the redshift distribution obtained for 574 quasars detected in early surveys showed peaks that were spaced periodically in log(1 + z). This spacing was relative to the strongest peak at z = 1.96 and is given by the relation

\begin{equation}
\Delta log(1+z)= 0.089. 
\end{equation}

Eqn 3 is hereafter referred to as the Karlsson relation. In Fig 2 the intrinsic redshifts predicted using this relation are plotted as filled circles at level -0.5. These values represent one of the two intrinsic redshift distributions considered below. Although some of the predicted redshifts coincide with the three regions of high density, several do not.

In a previous analysis of the distribution of QSOs around the Seyfert II galaxy NGC 1068 it was found that, when combined with the results of \citet{bur90}, it appears to be possible to fit QSO intrinsic redshifts by the relation

\begin{equation}
z_{iQ} = 0.62[N - 0.1M_{N}]
\end{equation}

where $N$ is an integer, and $M_{N}$ can have only certain well-defined values determined by a quantum number $n$ \citep{bel02a,bel02b,bel02c,bel02d,bel03a,bel03b}. For $N$ = 1, the predicted values are given by the relation

\begin{equation}
z_{iQ[N=1]} = (0.62 - 0.062n)
\end{equation}
where $n$ = 0,1,2,3..,9. This equation has a maximum redshift value of 0.62. For $N$ = 2 the predicted values are 0.31, 0.62, 0.868, 1.054, 1.178, and 1.24. For $N$ = 3 the predicted values are 0.558, 1.488, 1.798, and 1.86. The $N$ = 3 values are plotted in Fig 2 at level -1, and are indicated by the vertical lines. These redshifts correspond closely to the regions of high density in Fig 2. For $N$ = 4 the predicted intrinsic redshift values are 2.48, 2.418, and 1.178. For all higher $N$-states, z$_{iQ} > 3$ and therefore cannot be correlated with either of the observed redshift distributions considered here. 

Eqn 4 predicts the intrinsic redshifts that make up the second intrinsic redshift distribution considered below and is hereafter referred to as the z$_{iQ}$ relation. It was found previously that the QSOs near NGC 1068 were all in the $N$ = 2 state. Since all, or at least most, of the QSOs around NGC 6212 are assumed here to have been ejected from the same parent galaxy, it is assumed that they too will all be in the same $N$-state. From Fig 2, that state appears to be the $N$ = 3 state.

\begin{figure}
\hspace{-1.0cm}
\vspace{-1.0cm}
\epsscale{1.0}
\plotone{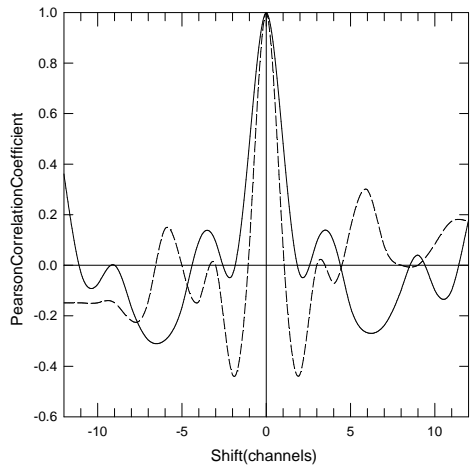}
\caption{\scriptsize{The zero shift channel gives the correlation coefficient r between (solid curve) the z$_{iQ}$ relation and itself, and (dashed curve)the Karlsson relation and itself.  \label{fig3}}}
\end{figure}

\begin{figure}
\hspace{-1.0cm}
\vspace{-1.0cm}
\epsscale{1.0}
\plotone{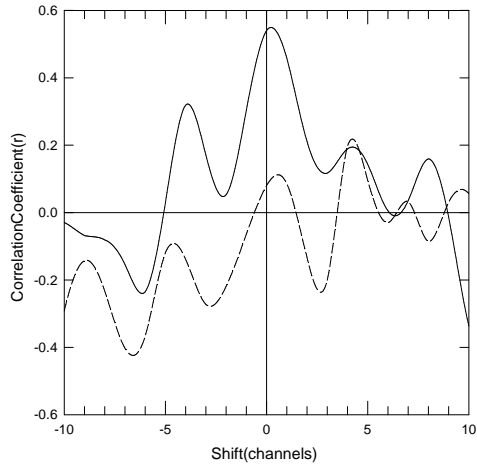}
\caption{\scriptsize{The zero shift channel gives the correlation coefficient r between the QSO redshift distribution and (solid curve) the intrinsic redshifts predicted from the z$_{iQ}$[$N$=3] relation and (dashed curve) the distribution predicted by the Karlsson relation.  \label{fig4}}}
\end{figure}

If the widths of the dense regions in the QSO redshift distribution are due to l-o-s ejection velocity smearing, as suggested above, clearly \em isolated \em redshift values cannot be expected to correlate well with individual intrinsic redshift values that are spaced apart by a distance less than, or the order of, this smearing, unless the objects have been ejected perpendicular to the l-o-s.
 
To check for a correlation we first divided the observed and predicted distributions into 28 redshift bins of width 0.1 in redshift. We then counted the number of sources in every 2 adjacent bins, centered every 0.1 in redshift. This resulted in redshift tables containing 28 channels for both the QSO distribution and each of the intrinsic redshift distributions. All tables were obtained with identical data sampling. It needs to be kept in mind that in addition to the intrinsic component the QSO redshifts contain ejection velocity components and a cosmological component, neither of which is present in the calculated intrinsic redshifts. The former is expected to smear out the intrinsic component present in the QSO redshifts, and the latter will shift the QSO redshifts (1 + z$_{o}$) up by a factor of (1 + z$_{c}$). The running mean over two bins was an attempt to introduce a small amount of smearing into the calculated intrinsic redshift distribution to more closely simulate the observed data. 

In order to get a feeling for what we might expect from the correlation analysis we first carried out an analysis between each intrinsic redshift distribution and itself. The correlation function used is given by

\begin{equation}
r = \frac{n(\Sigma XY) - (\Sigma X)(\Sigma Y)}{\sqrt{[n\Sigma X^2 -(\Sigma X)^2][n\Sigma Y^2 -(\Sigma Y)^2]}}
\end{equation}

which returns values between -1 and +1. Since there should be a perfect correlation between each data set and itself, the correlation coefficient must be +1.

The results are shown in channel 0 of the dashed curve in Fig 3 for the Karlsson relation, and in channel 0 of the solid curve for the z$_{iQ}$[$N$=3] relation. Values for the channels on either side of channel 0 were obtained by shifting one distribution relative to the other and recalculating the coefficient. This was done in order to obtain a baseline, and to provide correlation coefficients at nearby channels that would include the situations where one distribution has been shifted by an additional redshift component (eg. the shift due to the cosmological component in the QSO data). Note in Fig 3 that several additional positive and negative features are also present as expected in this type of analysis.

\begin{figure}
\hspace{-1.0cm}
\vspace{-1.0cm}
\epsscale{1.0}
\plotone{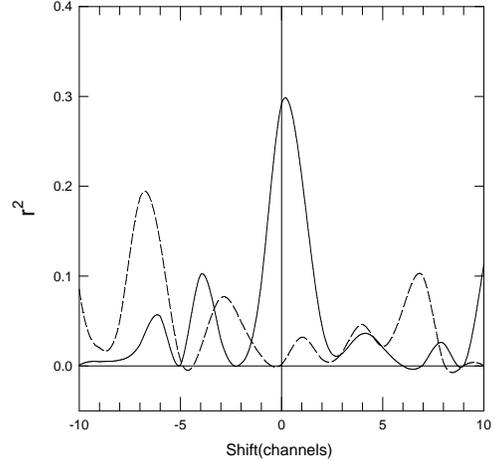}
\caption{\scriptsize{The zero shift channel gives r$^{2}$ between the QSO redshift distribution and (solid curve) the redshifts predicted from the z$_{iQ}$ relation and (dashed curve) the redshifts predicted by the Karlsson relation.  \label{fig5}}}
\end{figure}

\begin{figure}
\hspace{-1.0cm}
\vspace{-1.5cm}
\epsscale{1.0}
\plotone{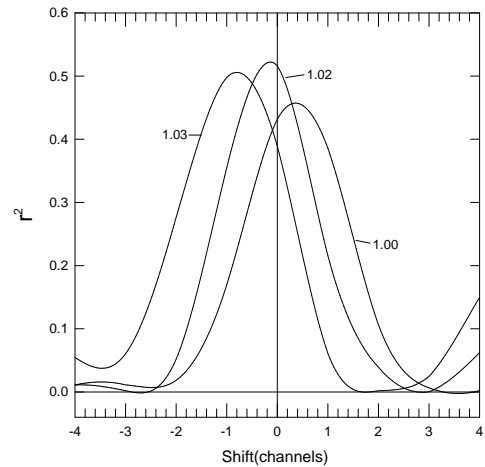}
\caption{\scriptsize{Correlation curves obtained for clumped sources after adjusting the z$_{iQ}$ intrinsic redshifts by the factor indicated.  \label{fig6}}}
\end{figure}

We next carried out a correlation calculation between the 28 samples in the QSO redshift distribution and the 28 samples in each of the two predicted intrinsic redshift distributions in turn. The results are presented in Fig 4 and show that there is no significant correlation between the QSO redshift distribution and the intrinsic redshift distribution predicted by the Karlsson relation. On the other hand, there appears to be evidence for a significant correlation (r $>$0.5) between the QSO distribution and the redshift distribution predicted by the z$_{iQ}$[$N$=3] intrinsic redshift relation.

We also carried out analyses in which there was no running mean taken. When the data were sampled in 28 bins with widths of 0.1 in redshift and no running mean, we obtained values of r = -0.018 and 0.378 for the Karlsson and z$_{iQ}$ relations respectively. When the data were sampled in 14 bins with widths of 0.2 in redshift we obtained r = 0.09 and 0.60 for these same two relations respectively. Both r-values obtained for the z$_{iQ}$ distribution are significant at the 5 percent level or better \citep[page 193]{whe68}.

In Fig 5 the squared correlation function (r$^{2}$) is shown for the distributions in Fig 4. Note that the peak is centered slightly above zero shift. This is discussed in more detail below. It is clear from Figs 4 and 5 that, although some of the redshifts predicted by the Karlsson relation may line up with those observed in the QSOs around NGC 6212, there is no significant correlation between the two.

There was also no evidence for a correlation between the observed redshift distribution and the z$_{iQ}$ relation for any $N$-value other than $N$ = 3.

\section{Is there evidence that the QSO redshifts contain a cosmological component?}

If the QSOs have been ejected from NGC 6212, their observed redshifts should contain a cosmological redshift component similar to that of NGC 6212. Seventy percent of the QSOs near NGC 6212 are clumped into three redshift groups whose smearing can be attributed to their l-o-s ejection velocities. These smeared-out clumps may then be a characteristic of the sources ejected from NGC 6212, suggesting that the sources not in broad clumps might be background sources. Thus, in order to examine the question of a cosmological component further, we removed the 12 sources that fell outside the clumps from the source list and ran the correlation analysis with the z$_{iQ}$[$N$=3] intrinsic redshifts again. The result is shown by the curve labeled 1.00 in Fig 6. The only significant difference between this curve and the corresponding one in Fig 5 is that the peak is higher. Both curves show a peak slightly above zero shift and this is assumed to be due the fact that the observed redshifts all contain a component due to the cosmological redshift of NGC 6212 that is not present in the calculated intrinsic redshifts. To investigate this further we adjusted each of the calculated intrinsic redshifts (1 + z$_{iQ}$) by factors of 1.02 and 1.03 and repeated the analysis. The results are shown by the two curves labeled 1.02 and 1.03 in Fig 6.

Several conclusions can be drawn from these results. First, the correlation coefficient has increased slightly, presumably since the correction is proportional to (1 + z$_{o}$) and the higher intrinsic redshifts are now a better fit to the QSO data. Second, the peak has moved through the zero shift position, which says that the cosmological adjustment is in the correct direction. Third, the best fit appears to be closer to z$_{c}$ = 0.02 than to 0.03, the observed redshift of NGC 6212. This could imply, as suggested earlier, that the observed redshift of NGC 6212 contains a small intrinsic component.

\section{Correlation between the 574 Karlsson quasar redshifts and those predicted by the two intrinsic redshift models} 

We also carried out a similar correlation analysis on the distribution of 574 quasar redshifts used initially by \citet{kar71,kar77} to derive the Karlsson relation. It is assumed here that the peaks in this distribution, shown in Fig 7, are visible because they have little, or no, l-o-s ejection velocity redshift component and only a very small cosmological component. Those that do contain a large Doppler component are assumed to make up the distribution curve denoted by the solid line. Since there is very litte Doppler smearing in the data associated with the peaks in Fig 7, the redshift distributions were obtained by simply counting sources in each 0.1 redshift interval over 28 intervals. Note also that since the redshifts in the Karlsson objects are not associated with a particular parent galaxy, objects in both $N$ = 2 and $N$ = 3 states are assumed to be present in the sample. The correlation results are shown in Fig 8 where the solid curve gives r$^{2}$ values between the Karlsson redshift distribution for 574 quasars and the distribution predicted by the z$_{iQ}$ relation (for both $N$ = 2 and 3). The dashed curve gives r$^{2}$ between the distribution of 574 quasars and the distribution given by the Karlsson log(1 + z) relation. Again, evidence for a significant correlation shows up only for the z$_{iQ}$ relation which has a corresponding correlation coefficient of r = 0.63. 

\section{Discussion}

Conventional techniques used to search for discrete intrinsic redshifts have mainly involved trying to find periodicities in raw redshift distributions \citep{arp02,bur68,bur90,kar71,kar77,tif97,haw02}. If the redshifts contain sizeable, and random, cosmological components, the intrinsic components will be smeared out. In fact, this technique is not even meaningful in intrinsic redshift models in which the entire redshift is assumed to be quantized, if there are large peculiar velocities present. Before meaningful observations can be made all Doppler components must be accounted for, as was attempted for those sources near NGC 1068 \citep{bel02c}. In the present paper it is possible to account for the cosmological component if the objects are at the distance of NGC 6212. There is also evidence, as we have seen above, that the ejection velocities of the QSOs near NGC 6212 are not large enough to completely mask the intrinsic redshifts predicted by the z$_{iQ}$[$N$=3] relation.

\begin{figure}
\hspace{-1.0cm}
\vspace{-1.0cm}
\epsscale{1.0}
\plotone{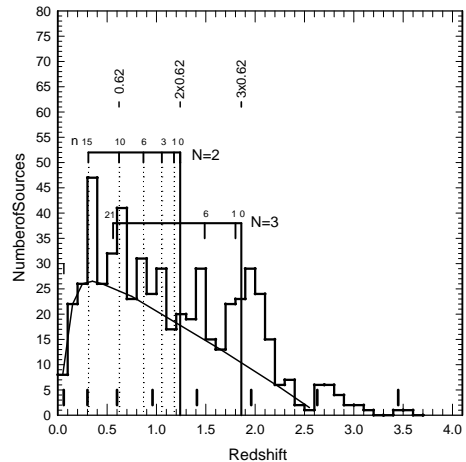}
\caption{\scriptsize{Redshift distribution for quasar data discussed by \citet{kar71,kar77}. The locations of the Karlsson intrinsic redshifts are indicated at the bottom and the z$_{iQ}$ redshifts at the top.  \label{fig7}}}
\end{figure}

\begin{figure}
\hspace{-1.0cm}
\vspace{-1.0cm}
\epsscale{1.0}
\plotone{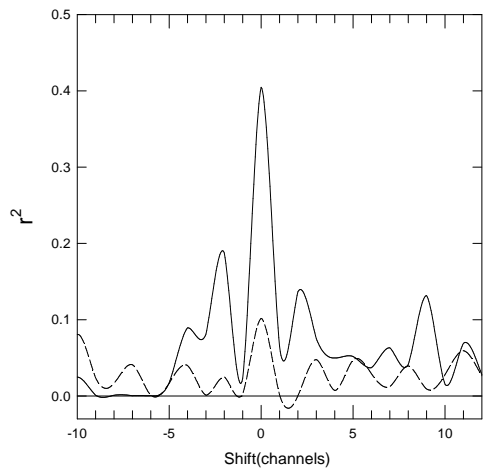}
\caption{\scriptsize{Squared Correlation coefficient between the Karlsson redshift distribution and, (solid) redshifts predicted by the z$_{iQ}$ relation for $N$ = 2 and 3, and, (dashed) redshifts predicted by the Karlsson relation. No running means used.  \label{fig8}}}
\end{figure}

We have found that the intrinsic redshift distribution predicted by the Karlsson relation is \em not \em correlated with either of the two observed redshift distributions. This relation also falls short when it comes to explaining many of the other quantized intrinsic redshift observations claimed. It does not predict the distribution of redshifts between z = 0.06 and z =0.62 reported by \citet{bur90}. Nor does it link the quantized redshifts found in quasars to the quantized 'velocities' found by \citet{tif96,tif97} to be present in galaxies.

On the other hand, the intrinsic redshift distribution predicted by the z$_{iQ}$ relation given by eqn 4 has been found here to show a significant correlation with both observed redshift distributions. This relation (for $N$ = 1) also predicts the 0.062 periodicity between z = 0.06 and z = 0.62 \citep{bur90}. It has also been shown \citep{bel02d} that there is a direct link between the quasar intrinsic redshifts predicted by eqn 4 and the "velocity" periodicities reported in galaxies by Tifft.
  
Recently \citet{fos02} have reported that they have found a BL Lac object (z = 0.43) lying in the direction of NGC 4698 (z = 0.0033). Although this object is assumed to be a background object by these investigators, if it were actually located in the galaxy it would have to have an intrinsic redshift near z = 0.43, assuming that its l-o-s ejection velocity is small. Since this intrinsic redshift value corresponds closely to the intrinsic redshift predicted by eqn 5 for $n$ = 3, or z$_{\rm iQ}[1,3] = 0.434$ \citep[Table 2]{bel02d}, we suggest that the possibility that this object is associated with NGC 4698 cannot yet be ruled out. The Karlsson intrinsic redshift relation does \em not \em predict this intrinsic redshift value.  

Finally, we make the following two comments. Since the distribution of 574 quasar redshifts in Fig 7 was the distribution that was used initially to define the Karlsson relation, it is somewhat surprising that it does not show a more significant correlation coefficient. On the other hand, since the z$_{iQ}$ relation was defined using a completely independent data sample, the correlation analysis carried out here between the z$_{iQ}$ redshifts and the Karlsson distribution in Fig 7 can be considered as an independent test of the z$_{iQ}$ relation. It appears to have passed this test successfully.

\section{Conclusion}

We have carried out correlation analyses on two different observed redshift distributions, comparing them to the intrinsic redshift distributions predicted by the Karlsson intrinsic redshift relation, and by the z$_{iQ}$ = 0.62[$N$ - 0.1$M_{N}$] relation.  No significant correlation is found between the Karlsson relation and either of the observed redshift distributions. However, there does appear to be a significant correlation between the redshift distribution observed for the QSOs near NGC 6212 and the redshift distribution predicted by the z$_{iQ}$ relation, for $N$ = 3. No evidence of a correlation was found for $N$-values of 1, 2, 4, or higher. A significant correlation is also found between the distribution of 574 quasar redshifts discussed by Karlsson and the redshifts predicted by the z$_{iQ}$ relation (for $N$ = 2 and 3). For the QSOs near NGC 6212 there is also evidence that there is indeed a redshift component due to the cosmological distance of that galaxy present in the data. However, the optimum value appears to be closer to z$_{c}$ = 0.02 suggesting that the observed redshift of NGC 6212 may also contain a small intrinsic redshift component. We conclude on the basis of these results, and the points discussed above, that although the Karlsson relation may predict some values that align with peaks in the observed redshift distributions, there is little evidence to suggest that this equation is related to, or has any bearing on, the production of intrinsic redshifts in quasars.


\end{document}